# Grounded Graphene-Based Nano-Waveguide with Chiral Cover and Substrate: New Theoretical Investigation

**Mohammad Bagher Heydari [1,*], Mohammad Hashem Vadjed Samiei [1]**

[1,*] School of Electrical Engineering, Iran University of Science and Technology (IUST), Tehran, Iran

[*] Corresponding author: mohammadbagher.heydari@alum.sharif.edu ; heydari.sharif@gmail.com

**Abstract:** In this paper, we propose a new analytical model for grounded chiral slab waveguides covered with graphene sheets. The general waveguide is constructed of a graphene sheet sandwiched between two different chiral layers (as substrate and cover layers), each one has the permittivity, the permeability and the chirality of $\varepsilon_c, \mu_c, \gamma_c$, respectively. The substrate is supposed to be a perfect electromagnetic conductor (PEMC), which is able to easily convert to a perfect electric conductor (PEC) or a perfect magnetic conductor (PMC). A novel matrix representation is obtained for the general structure to find its dispersion relation and other propagating parameters. To show the richness of the proposed waveguide, two new structures have been introduced and investigated. It has been shown that the modal properties of these exemplary structures are tunable by changing the chemical potential of the graphene and the chirality. The proposed general structure and its analytical model can be utilized for designing tunable plasmonic devices in THz frequencies.

## 1. Introduction

Nowadays, graphene plasmonics is a new emerging science, which studies the excitation of surface plasmon polaritons (SPPs) on the graphene [1-3]. Due to some fascinating features of these SPPs, such as high mode confinement, a large variety of graphene-based devices have been reported at THz frequencies, such as waveguides [4-14], isolator [15], circulator [16, 17], coupler [18], resonator [19], antennas [20-22], filter [23], Radar Cross-Section (RCS) reduction-based devices [24-26], and graphene-based medical components [27-33]. It should be noted that noble metals support SPPs at the near-infrared and visible frequencies [19, 34-43].

Chiral materials, one of the interesting materials in chemistry science with many remarkable features such as optical activity and circular dichroism, have opened a new area for researchers to design and report innovative devices such as cloak [44], polarization rotator [45] and microstrip antenna [46]. The hybridization of graphene and chiral medium in a THz device is very fascinating because the plasmonic features of the proposed device can be controlled and adjusted simultaneously by changing the chirality and the chemical potential of the graphene [5, 47, 48]. For instance, the modal properties of a graphene-chiral interface have been discussed and studied in [47]. The authors in [5] have been investigated the effect of various factors on the plasmonic modes of a graphene-chiral-graphene slab waveguide.

In this paper, we aim to analyze and consider the plasmonic features of grounded chiral slab waveguides covered with graphene sheets. To the authors' knowledge, the general proposed structure and its analytical model has not been published in any research article. Our presented waveguide is a generalization of all types of hybrid graphene-chiral grounded slab waveguides. In our general structure, a graphene sheet has been sandwiched between two chiral layers, as cover and substrate layers. The ground is a PEMC boundary condition in the general form, which can be easily transformed into PEC or PMC boundaries. In this generalized structure, one side of the proposed waveguide is always grounded by a PEMC layer and the other side is open. It should be noticed that our presented study considers only chiral materials with real-valued chirality and the study of generalized bi-anisotropic materials with the complex-valued tensors (i.e. epsilon, mu, and gamma) is complicated and is outside the scope of this article. The general waveguide supports hybrid plasmonic modes, which are split into two kinds of modes, called "Higher modes" and "Lower modes" in this paper. Indeed, chiral materials are responsible for generating these modes.

The paper is organized as follows. Section 2 introduces the general waveguide and proposes its analytical model. New matrix representation is obtained in this section that its determinant achieves the propagation constant of the general structure. Section 3 explains how the dispersion relation can be computed for a specific case of the general structure and also review some numerical methods for solving nonlinear equations. In section 4, two new graphene-based waveguides with chiral substrates are studied to show the richness of our proposed structure. The first waveguide is composed of a graphene sheet located on the grounded chiral slab. The ground plane in this structure is supposed to be PEC. We have achieved high mode confinement ($n_{eff} = 28$) at the frequency of 20 THz for it. The second case is a graphene sheet placed on the chiral slab backed by a PEMC layer. The hybridization of graphene with chiral materials allows the designer to tune and control SPPs via the chirality and the chemical potential. Finally, section 5 concludes the article.

## 2. The Proposed Structure and its Analytical Model

In Fig. 1, the configuration of the proposed waveguide has been illustrated. In this structure, a graphene sheet has been sandwiched between two chiral layers (as cover and substrate layers) and it has isotropic conductivity, which is described by Kubo's formula [49]:

$$\sigma(\omega, \mu_g, \Gamma, T) = \frac{-je^2}{4\pi\hbar} Ln\left[\frac{2|\mu_g| - (\omega - j2\Gamma)\hbar}{2|\mu_g| + (\omega - j2\Gamma)\hbar}\right] + \frac{-je^2 K_B T}{\pi\hbar^2(\omega - j2\Gamma)}\left[\frac{\mu_g}{K_B T} + 2Ln\left(1 + e^{-\mu_g/K_B T}\right)\right] \quad (1)$$

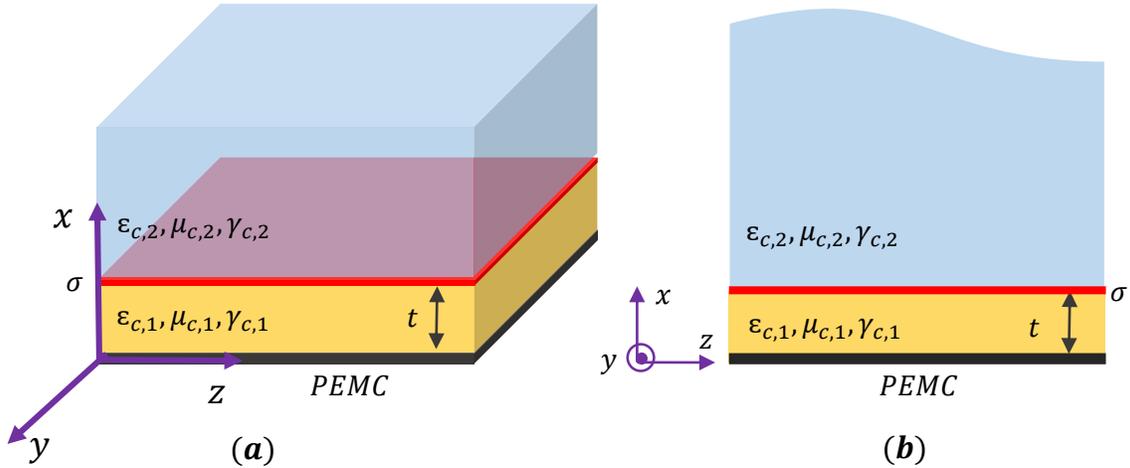

**Fig. 1.** The configuration of the proposed structure: **(a)** the 3D schematic, **(b)** the cross-section. The cover and substrate are assumed to be different chiral layers.

In the above relation, $\hbar$ is the reduced Planck's constant, $K_B$ is Boltzmann's constant, ω is radian frequency, $e$ is the electron charge, $\Gamma$ is the phenomenological electron scattering rate ($\Gamma = 1/\tau$, where $\tau$ is the relaxation time), $T$ is the temperature, and $\mu_g$ is the chemical potential which can be altered by chemical doping or electrostatic bias [49]. Moreover, each chiral layer satisfies the following relations [50]:

$$\boldsymbol{D}_N = \varepsilon_{c,N}\boldsymbol{E}_N - j\gamma_{c,N}\boldsymbol{B}_N \tag{2}$$

$$\boldsymbol{H}_N = -j\gamma_{c,N}\boldsymbol{E}_N + \frac{\boldsymbol{B}_N}{\mu_{c,N}} \tag{3}$$

Where $\varepsilon_{c,N}, \mu_{c,N}, \gamma_{c,N}$ are the permittivity, the permeability, and the chirality of that layer ($N$ shows the index of each layer, $N=1, 2$), respectively. Furthermore, $\varepsilon_0, \mu_0$ are the permittivity and permeability of free space, respectively. In our proposed structure, the ground plane is a PEMC layer with the following boundary condition [51]:

$$\hat{n} \times (\bar{H} + M\bar{E}) = 0 \tag{4}$$

In relation (4), $M$ is a PEMC admittance that can be transformed into PEC ($M \to \pm\infty$) and PMC ($M = 0$) boundaries.

To obtain the analytical model for our waveguide, we write the electromagnetic fields inside the chiral medium as the superposition of left-handed circularly polarized (LCP) and right-handed circularly polarized (RCP) waves [50]. Here, we assume that the electromagnetic waves propagate in the z-direction ($e^{j\beta z - i\omega t}$) and they have no variations in the y-direction ($\frac{\partial}{\partial y} = 0$). Therefore, the Helmholtz's equation for the $N$-th layer ($N=1, 2$) is [52]:

$$\frac{d^2\Psi_{R,N}}{dx^2} + q_{R,N}^2 \Psi_{R,N} = 0 \tag{5}$$

$$\frac{d^2\Psi_{L,N}}{dx^2} + q_{L,N}^2 \Psi_{L,N} = 0 \tag{6}$$

Where

$$q_{R,N} = \sqrt{\beta^2 - k_{R,N}^2} \tag{7}$$

$$q_{L,N} = \sqrt{\beta^2 - k_{L,N}^2} \tag{8}$$

In (7)-(8), $k_{R,N}, k_{L,N}$ are the propagation constants of RCP and LCP waves [50]:

$$k_{R,N} = \omega(n_{c,N} + \gamma_{c,N}) \tag{9}$$

$$k_{L,N} = \omega(n_{c,N} - \gamma_{c,N}) \tag{10}$$

Where

$$n_{c,N} = \sqrt{\frac{\mu_{c,N}\varepsilon_{c,N}}{\mu_0\varepsilon_0}} \tag{11}$$

is the refractive index of the $N$-th chiral layer [50]. It should be mentioned that the impedance of the $N$-th chiral layer has been defined as follows [50]:

$$\eta_{c,N} = \sqrt{\frac{\mu_{c,N}}{\varepsilon_{c,N}}} \tag{12}$$

Hence, the electromagnetic fields inside the chiral medium can be expressed as [53]:

$$\Psi_{R,N} = E_{z,N} + j\eta_{c,N}H_{z,N} \tag{13}$$

$$\Psi_{L,N} = E_{z,N} - j\eta_{c,N}H_{z,N} \tag{14}$$

Which results in [53]:

$$E_{z,N} = \frac{\Psi_{R,N} + \Psi_{L,N}}{2} \tag{15}$$

$$H_{z,N} = \frac{\Psi_{R,N} - \Psi_{L,N}}{2j\eta_{c,N}} \tag{16}$$

The transverse components of electric and magnetic fields are obtained by utilizing Maxwell's equations:

$$\begin{pmatrix} E_{x,N} \\ E_{y,N} \end{pmatrix} = -\frac{1}{2} \overline{\overline{Q}}_N^E \cdot \frac{\partial \Phi_N}{\partial x} \tag{17}$$

$$\begin{pmatrix} H_{x,N} \\ H_{y,N} \end{pmatrix} = -\frac{1}{2\eta_{c,N}} \overline{\overline{Q}}_N^H \cdot \frac{\partial \Phi_N}{\partial x} \tag{18}$$

Where the following definitions have been used in the relations (17)-(18):

$$\overline{\overline{Q}}_N^E = \begin{pmatrix} j\beta & j\beta \\ k_R & k_L \end{pmatrix} \tag{19}$$

$$\overline{\overline{Q}}_N^H = \begin{pmatrix} \beta & -\beta \\ -jk_R & jk_L \end{pmatrix} \tag{20}$$

$$\Phi_N = \begin{pmatrix} \dfrac{\Psi_{R,N}}{q_{R,N}^2} \\ \dfrac{\Psi_{L,N}}{q_{L,N}^2} \end{pmatrix} \tag{21}$$

Let us suppose that the propagation constants of RCP and LCP waves are:

$$q_R = \begin{cases} q_{R,1} & 0 < x < t \\ j q_{R,2} & x > t \end{cases} \tag{22}$$

$$q_L = \begin{cases} q_{L,1} & 0 < x < t \\ j q_{L,2} & x > t \end{cases} \tag{23}$$

Now, the Helmholtz's relations are written as:

$$\Psi_R = \begin{cases} A_R \cos(q_{R,1} x) + B_R \sin(q_{R,1} x) & 0 < x < t \\ C_R \exp(-q_{R,2} x) & x > t \end{cases} \tag{24}$$

$$\Psi_L = \begin{cases} A_L \cos(q_{L,1} x) + B_L \sin(q_{L,1} x) & 0 < x < t \\ C_L \exp(-q_{L,2} x) & x > t \end{cases} \tag{25}$$

Where $A_R, B_R, C_R, A_L, B_L, C_L$ are the unknown coefficients and should be determined by applying boundary conditions. By using relation (4), boundary conditions at $x = 0$ are obtained as [51]:

$$H_{y,1} + M E_{y,1} = 0 \; , \; H_{z,1} + M E_{z,1} = 0 \tag{26}$$

At $x = t$, the boundary conditions are written as [54]:

$$E_{z,1} = E_{z,2} = E_z \; , \; E_{y,1} = E_{y,2} = E_y \tag{27}$$

$$H_{z,2} - H_{z,1} = \sigma E_y \; , \; H_{y,2} - H_{y,1} = -\sigma E_z \tag{28}$$

By applying boundary conditions expressed in (26)-(28), the final matrix representation is achieved:

$$\overline{\overline{S}}_{6,6} \cdot \begin{pmatrix} A_R \\ A_L \\ B_R \\ B_L \\ C_R \\ C_L \end{pmatrix}_{6,1} = \begin{pmatrix} 0 \\ 0 \\ 0 \\ 0 \\ 0 \\ 0 \end{pmatrix}_{6,1} \tag{29}$$

In (29), the matrix $\overline{\overline{S}}$ is:

$$\bar{\bar{S}} = \begin{pmatrix}
0 & 0 & \dfrac{k_{R,1}}{q_{R,1}}\left(\dfrac{j}{\eta_{c,1}}-M\right) & -\dfrac{k_{L,1}}{q_{L,1}}\left(\dfrac{j}{\eta_{c,1}}+M\right) & 0 & 0 \\
\left(\dfrac{1}{j\eta_{c,1}}+M\right) & \left(M-\dfrac{1}{j\eta_{c,1}}\right) & 0 & 0 & 0 & 0 \\
\cos(q_{R,1}t) & \cos(q_{L,1}t) & \sin(q_{R,1}t) & \sin(q_{L,1}t) & -\exp(-q_{R,2}t) & -\exp(-q_{L,2}t) \\
-\dfrac{k_{R,1}\sin(q_{R,1}t)}{q_{R,1}} & -\dfrac{k_{L,1}\sin(q_{L,1}t)}{q_{L,1}} & \dfrac{k_{R,1}\cos(q_{R,1}t)}{q_{R,1}} & \dfrac{k_{L,1}\cos(q_{L,1}t)}{q_{L,1}} & \dfrac{k_{R,2}\exp(-q_{R,2}t)}{q_{R,2}} & \dfrac{k_{L,2}\exp(-q_{L,2}t)}{q_{L,2}} \\
-\dfrac{\cos(q_{R,1}t)}{j\eta_{c,1}} & \dfrac{\cos(q_{L,1}t)}{j\eta_{c,1}} & -\dfrac{\sin(q_{R,1}t)}{j\eta_{c,1}} & \dfrac{\sin(q_{L,1}t)}{j\eta_{c,1}} & \left(\dfrac{1}{j\eta_{c,2}}-\dfrac{k_{R,2}\sigma}{q_{R,2}}\right)\times\exp(-q_{R,2}t) & -\left(\dfrac{1}{j\eta_{c,2}}+\dfrac{k_{L,2}\sigma}{q_{L,2}}\right)\times\exp(-q_{L,2}t) \\
j\dfrac{k_{R,1}\sin(q_{R,1}t)}{\eta_{c,1}q_{R,1}} & -j\dfrac{k_{L,1}\sin(q_{L,1}t)}{\eta_{c,1}q_{L,1}} & -j\dfrac{k_{R,1}\cos(q_{R,1}t)}{\eta_{c,1}q_{R,1}} & j\dfrac{k_{L,1}\cos(q_{L,1}t)}{\eta_{c,1}q_{L,1}} & \left(\sigma-\dfrac{j}{\eta_{c,2}q_{R,2}}\right)\times\exp(-q_{R,2}t) & \left(\sigma+\dfrac{j}{\eta_{c,2}q_{L,2}}\right)\times\exp(-q_{L,2}t)
\end{pmatrix}$$

(30)

Now, by setting $\det(\bar{\bar{S}}) = 0$, the dispersion relation (or the propagation constant, $\beta$) is found. Then, achieving plasmonic parameters such as the effective index ($n_{eff} = Re[\beta]/k_0$) and the propagation loss ($L_{Prop} = 1/(2Im[\beta])$) is straightforward.

## 3. Methods

In this section, we briefly study the calculation of the dispersion relations and also review numerical techniques for solving them. First, we will compute the dispersion relation for a specific case and second, we will consider the numerical solution of the relation (30) in MATLAB software.

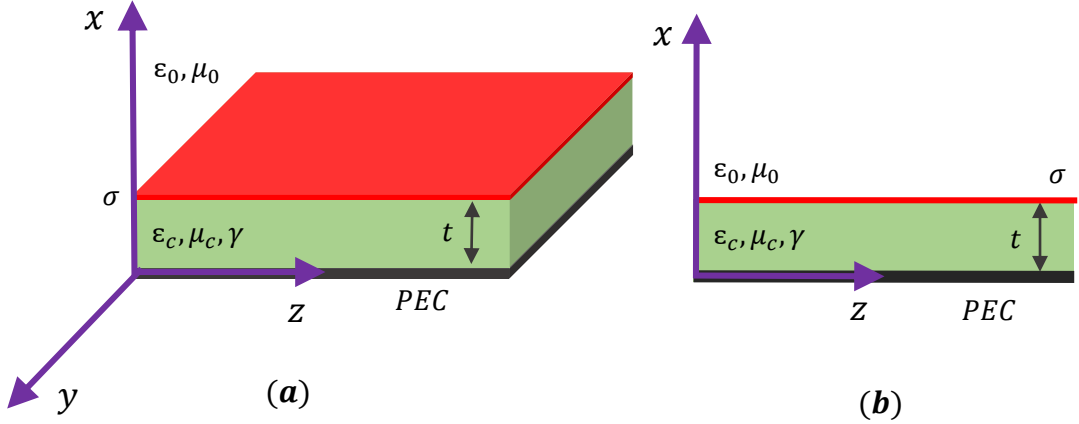

**Fig. 2.** The graphene-based waveguide with a chiral substrate backed by a PEC layer: **(a)** The 3D schematic, **(b)** The cross-section of the structure in the *z-x* plane.

Consider the waveguide depicted in Fig. 2, where a graphene sheet has been placed on the grounded chiral layer. The ground plane is supposed to be PEC. Hence, we will utilize the mathematical procedures of the previous section for $M \to \infty$. Consider the propagation constants of RCP and LCP waves for this structure:

$$q_R = \begin{cases} q_R & 0 < x < t \\ jq_0 & x > t \end{cases} \tag{31}$$

$$q_L = \begin{cases} q_L & 0 < x < t \\ jq_0 & x > t \end{cases} \tag{32}$$

Now, the Helmholtz's relations are written as:

$$\Psi_R = \begin{cases} A_R \cos(q_R x) + B_R \sin(q_R x) & 0 < x < t \\ C_R \exp(-q_0 x) & x > t \end{cases} \tag{33}$$

$$\Psi_L = \begin{cases} A_L \cos(q_L x) + B_L \sin(q_L x) & 0 < x < t \\ C_L \exp(-q_0 x) & x > t \end{cases} \tag{34}$$

Where $A_R, B_R, C_R, A_L, B_L, C_L$ are the unknown coefficients and should be determined by applying boundary conditions. By using the relations (26)-(28) for $M \to \infty$, the closed-form of the dispersion relation for the structure is derived:

$$\exp(-2jq_0 t) \times \begin{bmatrix} P_1 \cos^2(q_R t) + P_2 \cos(q_R t).\sin(q_R t) + P_3 \cos(q_R t).\cos(q_L t) + P_4 \sin^2(q_R t) \\ + P_5 \sin(q_R t).\sin(q_L t) + P_6 \cos(q_L t).\sin(q_R t) + P_7 \sin^2(q_L t) \end{bmatrix} = 1 \tag{35}$$

Where

$$P_1 = 2\eta_c \sigma k_0^2 k_R q_R k_L q_L - 2k_0 k_R k_L q_0 q_R q_L + 2k_R^2 q_0^2 k_L q_L - \eta_c \sigma q_0^2 k_R q_R k_L q_L \tag{36}$$

$$P_2 = 2\eta_c \sigma k_0^2 k_R q_R k_L q_L - (2+2j).k_0 k_R k_L q_0 q_R q_L + 2k_R^2 q_0^2 k_L q_L - \eta_c \sigma q_0^2 k_R q_R k_L q_L \\ -(4j).k_R q_L^2 q_R + (2j).k_R^2 q_L^2 k_0 q_0 \tag{37}$$

$$P_3 = 2\eta_c \sigma k_0^2 k_R q_R k_L q_L + 2k_0 k_R k_L q_0 q_R q_L + 2k_L^2 q_0^2 k_R q_R - \eta_c \sigma q_0^2 k_R q_R k_L q_L \tag{38}$$

$$P_4 = 2\eta_c \sigma k_0^2 k_R q_R k_L q_L - 2k_0 k_R k_L q_0 q_R q_L - 2k_R^2 q_0^2 k_L q_L - \eta_c \sigma q_0^2 k_R q_R k_L q_L \tag{39}$$

$$P_5 = 2\eta_c \sigma k_0^2 k_L^2 q_R^2 + 2\eta_c \sigma k_0^2 k_R^2 q_L^2 - 2q_0 q_R^2 k_0 k_L^2 + 2k_0 k_R^2 q_0 q_L^2 - 2k_L^2 k_R q_0^2 q_R - \eta_c \sigma k_L^2 q_0^2 q_R^2 \\ + 2k_R^2 k_L q_0^2 q_L - \eta_c \sigma k_R^2 q_0^2 q_L^2 \tag{40}$$

$$P_6 = 2k_0^2 k_R q_R k_L q_L + (4j).q_L k_L q_R^2 + (2j).k_0 k_L^2 q_0 q_R^2 + (2+2j).k_0 k_R k_L q_0 q_R q_L \\ + 2k_R k_L^2 q_R q_0^2 - \eta_c \sigma k_R q_R q_0^2 k_L q_L \tag{41}$$

$$P_7 = 2\eta_c \sigma k_0^2 k_R q_R k_L q_L + 2k_0 k_R k_L q_0 q_R q_L + 2k_L^2 q_0^2 k_R q_R - \eta_c \sigma q_0^2 k_R q_R k_L q_L \tag{42}$$

As seen in the above relations, the closed-form relation is so complicated and it is better to use numerical methods for solving the general matrix relation of (30). Some numerical software such as MATLAB can calculate the determinant of a matrix with an arbitrary size, by using "Cofactor expansion" [55]. Here, to obtain the determinant of matrix (30), one should define the frequency and the propagation constant as symbolic (by using command "syms" in MATLAB) and then utilize the command "det" to find the determinant of general matrix (defined in (30)) as a function of frequency and the propagation constant. Let us assume that the outcome of MATLAB software for the determinant of the relation (30) is:

$$\det\left(\overline{\overline{S_{6\times 6}}}\right) = f(\omega, \beta) \tag{43}$$

Now, to solve the nonlinear equation of $det(\bar{\bar{S}}) = 0$, numerical methods in mathematics such as Newton-Raphson [56] can be used, which exist in MATLAB software. Hence, one can utilize MATLAB software to solve the nonlinear equation of $det(\bar{\bar{S}}) = 0$.

## 4. Special Cases of the Proposed Structure: Results and Discussions

This section introduces and studies two new graphene-based waveguides incorporating chiral substrates. Without loss of generality, the cover layer in both structures is supposed to be air. The substrate is a chiral layer in both waveguides and it has different parameters in them. Also, the ground plane is PEC and PEMC boundary conditions in the first and second structures, respectively. Both waveguides support hybrid plasmonic modes, which are tunable via the chirality and the chemical potential of the graphene sheet. In chiral-based waveguides, hybrid modes are split into two "bifurcated modes", which is one of the interesting properties of these structures [50, 57, 58]. In this paper, we call them "Higher modes" and "Lower modes" that are relevant to high and low plasmon resonance frequencies, respectively. Bifurcated modes are hybrid modes with the same cut-off frequency and different propagation constants. In our studied structures, chiral materials are responsible for generating these modes. In the literature, mode bifurcation has been reported in some structures such as parallel plate waveguides [50], circular waveguides [58], rectangular waveguides [57] and diagonal anisotropic chiral waveguides [59]. The nature of bifurcating modes depends on the studied chiral structure. For instance, in circular waveguides, modes with $e^{-im\varphi}$ and $e^{im\varphi}$ ($m \neq 0$) azimuth variations have different phase constants (above the cut-off frequency) and the same cut-off frequency, which form bifurcated modes [58]. Our proposed, general structure and its specific cases (the first and the second structures) are a kind of parallel plate waveguides. In these structures, odd and the next even mode have the same cut-off frequency and thus they are a pair of bifurcated modes [50, 58]. The mathematical explanations of emerging bifurcated modes in parallel plate chiro-waveguides have been reported in [50].

In all the following analytical results, the temperature is $T = 300\ K$ and the relaxation time of the graphene is $\tau = 0.4\ ps$. The chemical potential of the graphene is assumed to be $\mu_g = 0.3\ eV$ unless otherwise stated. It is worthwhile to be mentioned that the relaxation time of the graphene depends on the phenomenological scattering rate and the quality of graphene [1, 60-62]. In what follows, we will investigate the modal properties of these waveguides.

*4.1 The First Structure: A Graphene-Based Waveguide with a Chiral Substrate Backed by a PEC Layer*

The configuration of the first structure was introduced in the previous section (see Fig. 2) and a closed-form equation was obtained for its dispersion relation. Here, we will consider and investigate the numerical results. In our simulations, the chiral substrate has a thickness of $t = 600\ nm$ and the effective index of $n_c = 1.3$. To indicate the tunability of plasmonic waves via chirality, two values of chirality (i.e. $\gamma = 0.0025, 0.0035\ \Omega^{-1}$) have been considered here. These values for the chirality and the index of the chiral medium depend upon the nature and structure of the organic, inorganic and biochemical molecules, which are reported in [63-69].

Fig. 3 represents the analytical results of the effective index and the propagation length. As mentioned before, two plasmonic modes (Higher and Lower modes) are appeared here due to the existence of the chiral substrate. One can observe from Fig. 3(a) that the effective index increases as the frequency increases. While the propagation length decreases with the frequency increment, as seen in Fig. 3(b). It happens because the imaginary part of the conductivity increases with the frequency increment and thus the propagation length decreases. Furthermore, it is seen from Fig. 3(b) that the propagation length of higher modes has larger values compared to the lower modes. A large value of the effective index, amounting to 28 e.g., is seen for the chirality of 0.0035 $\Omega^{-1}$ at the frequency of 20 THz for the lower mode.

In Fig. 4, we depict the effective index of higher mode for various chiral thicknesses ($t = 600, 650, 700$ nm). In this diagram, the chirality of the chiral substrate is 0.0025 $\Omega^{-1}$. This figure clearly indicates that the effective index increases as the thickness of the chiral slab decreases. It happens due to the extreme penetration of the electromagnetic fields inside the waveguide for thinner chiral thickness.

Fig. 5 illustrates the normalized profile distributions of $Re\,[E_x]$, $Re\,[E_y]$ and $Re\,[E_z]$. It can be seen from this figure that higher and lower modes have similar field distributions of $Re\,[E_z]$ and $Re\,[E_x]$. The field distribution of $Re\,[E_y]$ is the main difference between higher and lower modes, which can distinguish the different types of modes.

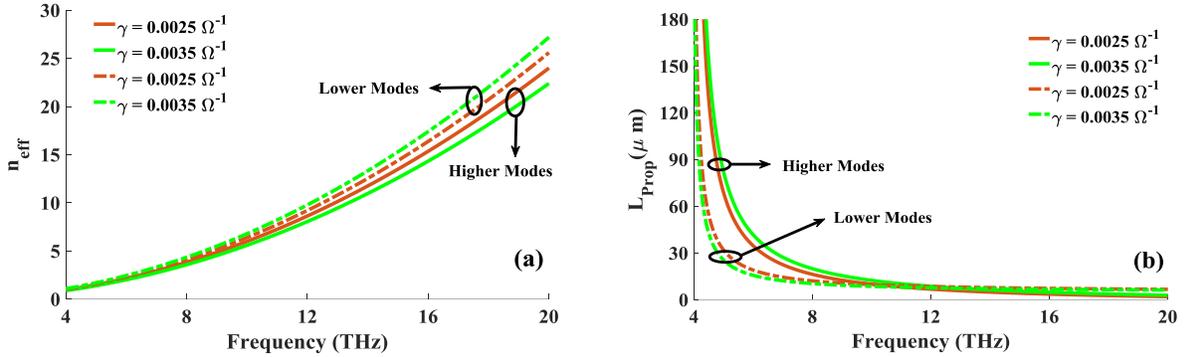

**Fig. 3.** Analytical results of modal properties for the first structure for various chirality values: **(a)** the effective index as a function of frequency for lower and higher hybrid modes, **(b)** the propagation length as a function of frequency for lower and higher hybrid modes. The chemical potential of the graphene sheet is 0.3 eV.

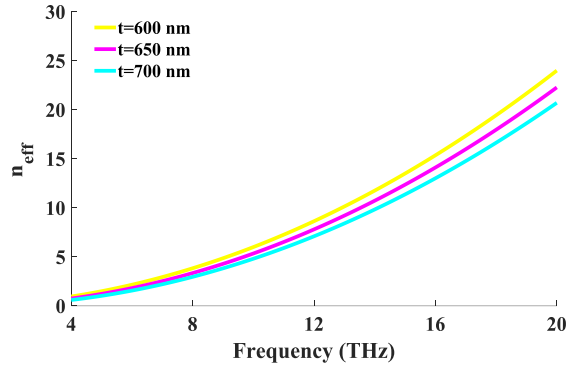

**Fig. 4.** The effective index of higher mode for the first structure for various chiral thicknesses ($t = 600, 650, 700$ nm). The chemical potential of the graphene sheet is 0.3 eV and the chirality of the chiral medium is 0.0025 $\Omega^{-1}$.

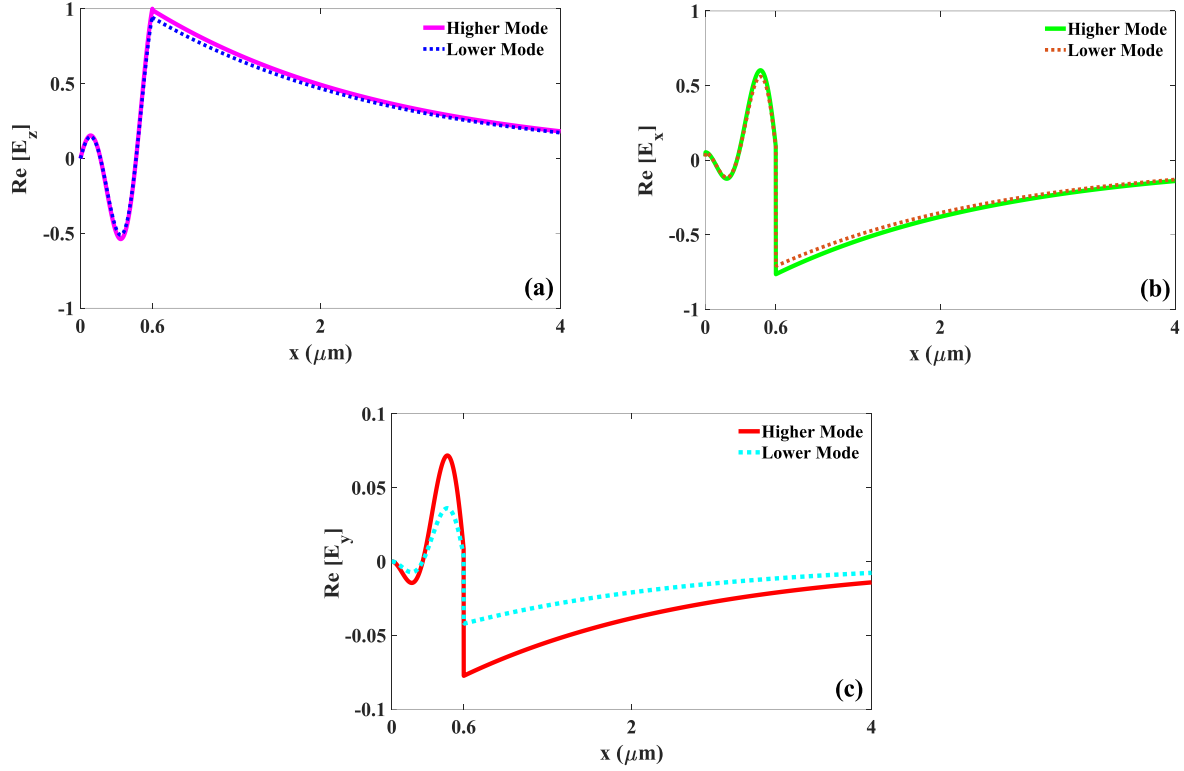

**Fig. 5.** Normalized field distributions of higher and lower modes: (a) Re [$E_z$], (b) Re [$E_x$], (c) Re [$E_y$], with $\mu_g$ = 0.3 eV, $\gamma$ = 0.0035 $\Omega^{-1}$ and f = 14 THz.

*4.2 The Second Structure: A Graphene-Based Waveguide with a Chiral Substrate Backed by a PEMC Layer*

The second structure, as seen in Fig. 6, is a graphene-based waveguide with a chiral substrate, where the ground plane is supposed to be a PEMC. In the second structure, the chiral slab has the thickness of $t = 750\ nm$ and its effective index is $n_c = 1.4$. The admittance parameter of PEMC is considered $M = 0.5$ unless otherwise stated. Similar to the first structure, two values of the chirality (here $\gamma = 0.003, 0.004\ \Omega^{-1}$) are considered for the chiral substrate to study the effect of chirality on the propagation properties.

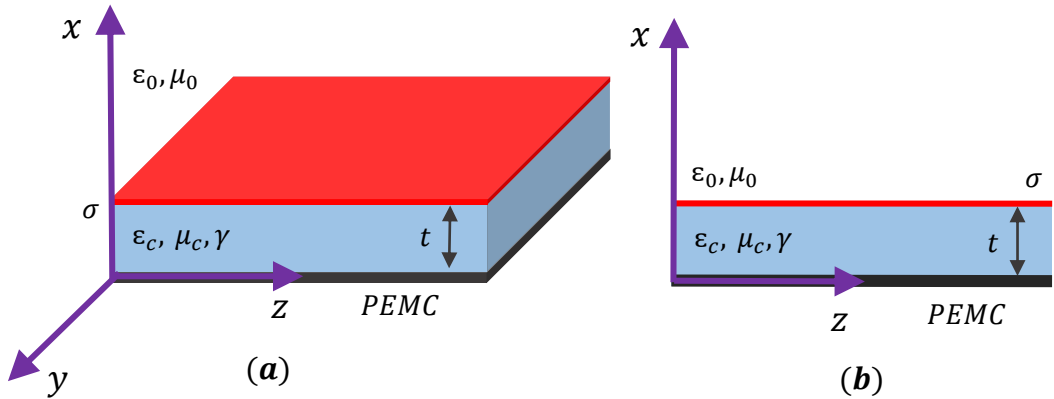

**Fig. 6.** The graphene-based waveguide with a chiral substrate backed by a PEMC layer: **(a)** The 3D schematic, **(b)** The cross-section of the structure in the *z-x* plane.

Fig. 7 illustrates the plasmonic features of the second structure as a function of the frequency. As seen in this figure, the effective index increases as the frequency increases while the propagation length decreases. Similar to the first structure, the propagation length reduces as the frequency increases because the imaginary part of the graphene conductivity (or losses) increases with the frequency increment. Moreover, the lower modes have larger values of the effective index compared to the higher modes.

Fig. 8 represents the dependence of the mode confinement on the chemical potential of the graphene. This figure has some remarkable features. First, it allows one to design the proposed structure at the desirable effective index. For instance, to achieve $n_{eff} = 8$, one should design the waveguide with the following parameters: $\gamma = 0.003 \, \Omega^{-1}, \mu_g = 0.2 \, eV$ or $\gamma = 0.004 \, \Omega^{-1}, \mu_g = 0.35 \, eV$. Second, the diagram determines the effective index of higher and lower modes at the specific chemical potential. Consider $\mu_g = 0.5 \, eV$. In this chemical doping, the effective index of higher and lower modes for the chirality of $\gamma = 0.004 \, \Omega^{-1}$ are $n_{eff} = 3, 6$, respectively. The last feature of this diagram is its ability for determining the propagation range of the plasmonic waves. For instance, hybrid modes do not propagate inside the waveguide for the chemical range of $\mu_g > 0.4 \, eV$ for the chirality of $\gamma = 0.003 \, \Omega^{-1}$. As seen in Fig. 8, the chirality loses its effect when the chemical potential of graphene increases, and thus the structure supports only one plasmonic mode (two propagating modes are converged). We called these points "cut-off points", which happens for the chemical potential of 0.4 eV (for the chirality of 0.003 $\Omega^{-1}$) and 0.7 eV (for the chirality of 0.004 $\Omega^{-1}$) for this studied structure at the frequency of 10 THz. At these points, two plasmonic modes are converted to radiation modes, i.e. the two lines collapse to a $n_{eff} = 1$ line.

The cut-off region for the structure of Fig. 6 can be obtained by supposing $k_{R,1} > k_{L,1}$ and setting $\beta = k_{R,1}, q_{R,1} = 0, q_{L,1} = \sqrt{k_{R,1}^2 - k_{L,1}^2}$. Fig. 9 shows the chiral cut-off value as a function of frequency for various chemical potentials. At a specific frequency (for instance, consider $f = 20 \, THz$), higher cut-off values are obtainable for low values of chemical potential.

As a final point, the effective indices of higher and lower modes have been depicted for various values of PEMC admittance (M). As seen in Fig. 10, the mode confinement increases with the admittance increment. Therefore, one way for achieving a desirable effective index is the design of the structure in the appropriate PEMC admittance.

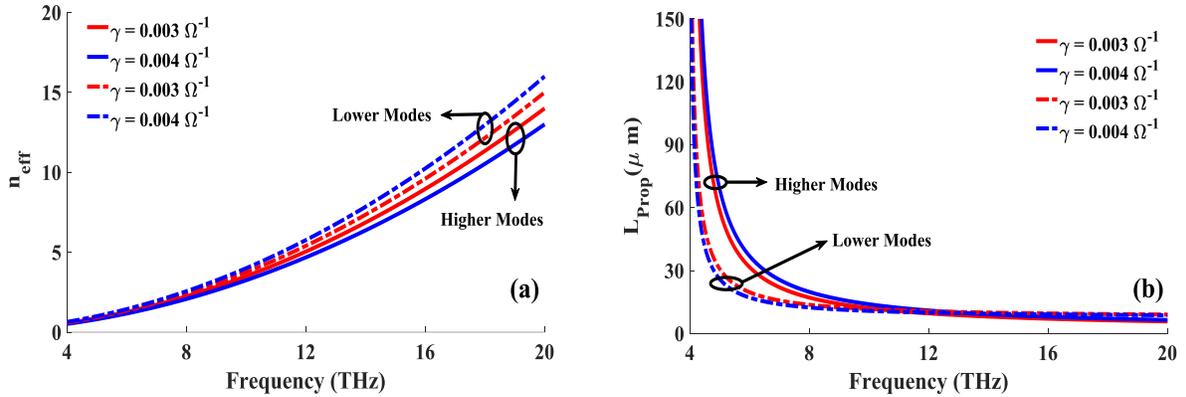

**Fig. 7.** Analytical results of modal properties for the second structure for various chirality values: **(a)** the effective index as a function of frequency for lower and higher hybrid modes, **(b)** the propagation length as a function of frequency for lower and higher hybrid modes. The chemical potential of the graphene sheet is 0.3 eV.

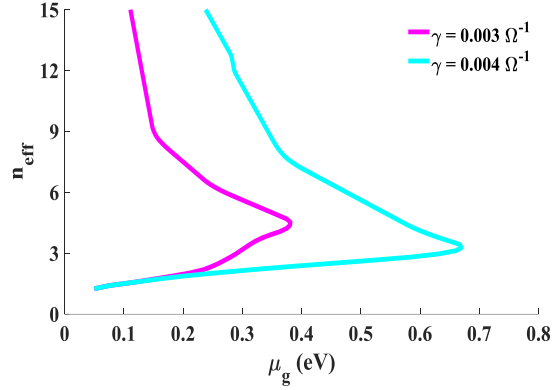

**Fig. 8.** The effective index of hybrid modes as a function of the chemical potential for the second waveguide at the frequency 10 THz.

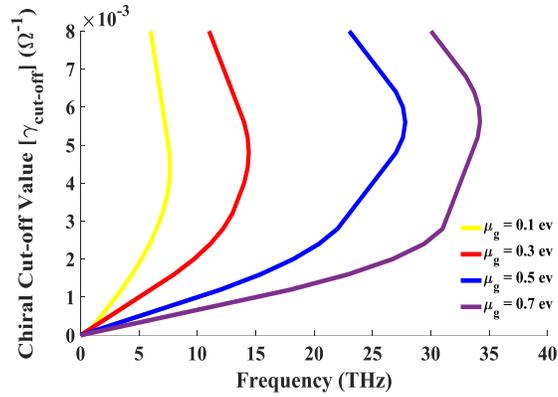

**Fig. 9.** Chiral cutoff values as a function of the frequency for various values of chemical potential.

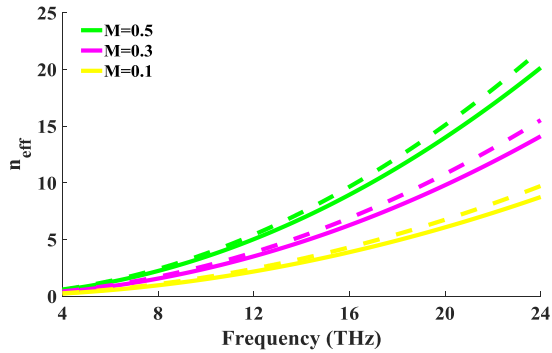

**Fig. 10.** The effective indices of higher and lower modes for the second structure for various admittances of PEMC ($M$ = 0.5, 0.3, 0.1). The chemical potential of the graphene sheet is 0.3 eV and the chirality of the chiral medium is 0.003 $\Omega^{-1}$. The dashed and solid lines represent lower and higher modes, respectively.

## 5. Conclusion

A new analytical model is presented for the proposed graphene-based waveguides in this article. The modal properties of plasmonic waves, propagating inside these waveguides, are adjustable by tuning the chirality and the chemical potential of the graphene. As special examples of the general rich structure, two novel waveguides have been proposed and considered. The first waveguide is composed of a graphene sheet deposited on the grounded chiral slab. The ground plane in this structure is supposed to be PEC. A large value of the effective index, amounting to 28 e.g., is obtained at the frequency of 20 THz. The second case is a graphene sheet placed on the chiral slab backed by a PEMC layer. It has been shown that the effective index of this structure can be changed by altering the admittance of the PEMC layer. The hybridization of graphene with chiral materials allows the designer to tune and control SPPs via the chirality and the chemical potential of the graphene. Our presented study can be utilized for designing new kinds of THz devices such as cloaks.